\documentclass[12pt]{iopart}

\usepackage{iopams}  
 \usepackage[pdftex]{graphicx}     
\begin{document}

\title[Persistence in PCPD]{Persistence as Order Parameter 
in Generalized Pair Contact Process with Diffusion}

\author{Maneesh B. Matte and Prashant M. Gade}
\address{PG Department of Physics, Rashtrasant Tukadoji Maharaj Nagpur
University, 
\\
Campus, Nagpur, 440 033, INDIA.}
\ead{prashant.m.gade@gmail.com}
\vspace{10pt}
\begin{indented}
\item[]February 2014
\end{indented}

\begin{abstract}
The question of universality class of pair contact process with 
diffusion (PCPD) is revisited with an alternative approach. We
study persistence in  Generalized Pair-Contact Process with 
diffusion (GPCPD) introduced by Noh and Park, (Phys.  Rev. E 
69,016122(2004)). This model allows us to
interpolate between directed percolation (DP) and PCPD universality classes.
We find that transition to nonzero persistence 
is at same parameter value as transition to zero 
number density.  We obtain finite size scaling and off-critical 
scaling collapse for persistence and find critical  exponents by
fitting phenomenological scaling laws to persistence.
While the dynamic scaling exponent $z$  varies  continuously
in GPCPD,  the 
correlation-time exponent $\nu_\parallel$ matches with directed 
percolation universality class.
\end{abstract}

\pacs{05.70.Fh,05.70.Ln}
%
\vspace{2pc}
\noindent{\it Keywords}: persistence, non-equilibrium phase transition, pair contact processes
%
\submitto{\JSTAT}
%
%
%

\section{Introduction}

Phase transitions in non-equilibrium systems have been investigated in 
detail in past two decades.  One of the earliest, most extensively 
studied and most commonly observed non-equilibrium transition is the
transition to an absorbing state. It is often described by Directed 
percolation (DP) universality class. 
Grassberger and  la Torre established  equivalence of Reggeon Field theory 
with Markov process. 
This was followed by Janssen and  Grassberger's  
conjecture that all transitions to
nondegenerate absorbing state in one-component 
systems without quenched disorder, with 
short range interactions in space and time, and in absence of multi-critical
points are in DP universality class. By now, we know several systems in this 
universality class.  
\cite{latorre,grassberger}

There are very few systems which are unambiguously shown
to be in an universality class different from DP. 
Systems such as branching and annihilating random walks with even
number of offsprings (BAWe)
\cite{jensen}, model A and B of probabilistic cellular
automata\cite{grassberger-ca},   
interacting monomer-dimer models \cite{park}  
and nonequilibrium kinetic Ising model  in presence of spin flip as
well as spin exchange dynamics \cite{odor} are found to be in directed
Ising (DI) universality
class\cite{park-3}. 
Voter universality class, parity conserving class (which has
same exponents as DI  in 1-d) and
absorbing phase transitions with conserved field
are such examples\cite{park-park}.
Compact directed percolation is another such class\cite{Mendes}.
Models such as
Manna model were considered to be in different universality
class. However, questions have been raised on the independent Manna 
class\cite{mohanty}.
Special boundary conditions, quenched disorder and anomalous diffusion
affect the exponents\cite{odor-rev}. 
In this work, we  study a process known as pair contact process with
diffusion (PCPD) which has been reported to be possibly in an independent
universality class using an alternative order parameter.

A model with infinite absorbing states called  pair contact process (PCP)
did attract lot of attention.   
In PCP, particles in 
two neighboring occupied sites  can annihilate each other with 
probability $p$ and produce an offspring with probability $1-p$.  Any state 
without pairs is an absorbing state.  Infinity of  such states are possible.
Despite not obeying sufficient conditions  for DP universality class,
this transition is considered to be in DP universality class\cite{lubeck-2}.

A variant of PCP known as pair contact process with diffusion (PCPD)
attracted even more attention. The universality 
class of this transition in 1-d has been a matter of long-standing 
debate and the issue is not yet settled. (for further generalization, 
see \cite{tcpd}.) Initially, it was seen as process in a distinctly
new universality class \cite{hinrich-01}.  But, several alternatives were
proposed in later studies on this model (See Sec. I of \cite{henkel}).   

One of the most plausible arguments
for PCPD being in DP class was given by Hinrichsen.
He argued that isolated particles can spread at most diffusively with
dynamic exponent 2. But most of the estimates of dynamic exponent of
PCPD yield value less than 2 implying a superdiffusive spread of
critical clusters. Hence the asymptotic behaviour should not be
dominated by diffusion of isolated particles which are effectively
frozen. If the isolated particles are frozen
PCPD reduces to PCP which
belongs to DP universality class\cite{physica}.

The model
shows very strong corrections to scaling. 
In such cases, it is argued that `very large scale and long-time
computations in future will help us to settle on an appropriate conclusion'.
This approach of longer and longer simulations on larger and larger
systems has been addressed to some extent
in recent decade and simulation time has reached
$10^9$ recent paper by Park\cite{park-14}. He has concluded
that on including corrections to scaling, the critical density decay exponent
is $0.173$ independent of diffusion constant. Thus this
model is incompatible with DP universality class.
Barkema and coworkers argue that the model is in same universality class as
DP by considering scaling of
ratio of pair density and particle density\cite{schram,smallenberg}
and assuming certain corrections to scaling.  
Due to strong corrections
to scaling, there is a debate even about critical density decay exponent.
On the theoretical side, this problem has resisted analysis 
and the latest work by Gredat {\it{et al}} in which PCPD has been studied
by nonperturbative functional renormalization group has
suggested that `critical behavior of PCPD can be either in DP or
a in a new (conjugated) class'\cite{gredat}.
The numerical studies on scaling of order parameter and
theoretical studies have not reached a conclusive end.
Hence, there is a need for an alternative approach. We work on a model
which reaches PCPD in certain limit and DP in another limit.

Critical density decay exponent $\delta$ for DP is known to be
0.1595 while that for PCPD has ranged from 0.17-0.275 \cite{smallenberg}
and the latest estimate is as low as 0.17 \cite{park-14}.
Thus the difference between exponents (if any) is very small. The value
of exponents is very small as well. It takes very long for system to 
settle into saturation state and time grows with system size. 
The static exponents which differ in magnitude much more
are difficult to compute.
For example, if we work on modest system size of 5 $\times$ $10^4$ 
lattice sites, it will need $10^{30}$ time steps before the number density
goes to zero at critical point.  Thus for a
small $\delta$, very long
simulations become necessary at large sizes.   In presence of
strong corrections to scaling, even these simulations
do not give reliable results.
Other parameters such as pair density decay in a similar manner and
they can help in correcting the corrections to scaling. 
We attempt an alternative order parameter which decays much faster.
Unfortunately, a larger decay exponent brings in its own 
difficulties such as larger relative error and difficulty in computing
static exponents.

In this work, we attempt 
to find indications of universality and scaling behavior
from short-time behavior check for
alternative `order parameter'.
Recently,  there have been attempts
to quench the system from high temperature 
to critical temperature, study its 
short time dynamics and analyze its relation with
the universality class\cite{Janssen,Huse}. There  have been
some successful attempts at obtaining critical exponents from  scaling 
behavior of persistence\cite{majumdar,Zheng,fuchs} which is again short-time
history dependent behavior. 
We study persistence in PCPD and  its generalized version 
from this point of view.

\section{Persistence}

Persistence is a generalization of first passage time for
spatially extended systems.  In spin systems, local persistence 
at  time $T$, $P(T)$ is the fraction of sites which did not 
change their initial spin state  at all times $t\leq T$. 
In Ising type systems,  
it may show power law decay in time at zero temperature and exponent 
is called persistence exponent.  In contact processes,
persistence may show power law at critical point 
and the exponent is called persistence exponent\cite{footnote2}.

The persistence exponents themselves 
are not universal and depend on delicate details of evolution.
Models falling in same universality class have the same 
persistence exponent in some cases, but not necessarily so.
For example, persistence exponent of widely different models such as 
1-d Ising model, coupled logistic maps 
and Sznajid model is $\frac{3}{8}$\cite{majumdar,stauffer,ggs}.
Similarly, it has been observed the persistence exponent in 1-D DP 
models such as Domany kinzel model\cite{koduvely}, Ziff-Gulari-Barshad
model \cite{albano}, site percolation\cite{fuchs,grass}  
and 1-d coupled circle maps \cite{menon} is same,
{\it i.e.} ${\frac{3}{2}}$ or very close to it.
In Ref. \cite{albano}, they carried out work on
2-d DP model and claimed that persistence exponent is superuniversal.
Now there is reasonable evidence that persistence
exponent is not even universal\cite{ali-gade}. Its value 
may be of interest from the viewpoint of detailed dynamics.
We argue that even if persistence exponent is not universal,
it can help in finding other universal exponents.

 This nontrivial exponent sheds  
further light on detailed dynamical nature of phase transitions 
\cite{majumdar}.  Furthermore, its scaling behavior also sheds  
light on other exponents in the system. Some
examples are as follows a)Fuchs {\it{et al}} studied 
1+1 dimensional contact process in DP class, and obtained a successful 
data collapse of persistence for finite size scaling as well as 
off-critical simulations using 1-d DP values of $\nu_\parallel$ 
and $z$\cite{fuchs}. 
b) The onset of spatiotemporal intermittency in 
coupled circle maps is known to be in DP universality class. 
Finite size scaling of persistence yields estimate of $z$ which matches
with 1-d DP value\cite{menon}. c) Hinrichsen and Koduvely 
studied persistence in Domany-Kinzel automata and found
that persistence obeys finite size scaling with same exponents as
DP\cite{koduvely}.  d) An evolutionary model of Prisoner's dilemma
on 2-d lattice is found to show a transition in DP universality class.  
Two variants of this model have been studied and $\nu_{\parallel}$ obtained
from off critical simulations of persistence in this model match with 2-d
DP values\cite{ali-gade}.  
e) Manoj and Ray obtained finite-size scaling 
collapse of  persistence in Glauber-Ising model in dimensions 1-4
\cite{manoj} at zero temperature.
Thus there are few cases  where transition is known and 
the scaling
collapse for persistence is obtained by using standard values.

We try this approach towards this problem and try to complete the
study. Local persistence in this model is not studied before. 
We study a model called GPCPD (Generalized Pair Contact
Process with Diffusion)  which smoothly interpolates between DP and 
PCPD with a parameter controlling memory strength \cite{noh}.
We compute the persistence exponent for this  model which is 
a new nontrivial critical exponent unrelated to other critical exponents. 

Apart from completing studies by determining persistence exponent, 
the studies can be used for validation of other critical exponents,
particularly when studies on order parameter have not been
conclusive.  As mentioned above, there have been
four prior cases in which persistence in 1-d and 2-d
DP models has been studied and finite size scaling and off-critical
scaling in these models give an estimate of $z$ and $\nu_\parallel$ which
agrees with standard DP models.  We observe a fairly convincing
scaling in GPCPD using this approach and determine scaling exponents in this
manner. 
In this model, for limiting case of DP, we 
indeed get scaling exponents close to those for DP.
Thus it is fair to expect that if PCPD were in DP class, 
DP scaling for persistence should prevail in finite size
scaling as well as off-critical simulations.
Our basic result is that while $z$ varies continuously
as reported by Noh and Park, 
the correlation-time exponent
$\nu_\parallel$ matches with DP for GPCPD.

\section{The model}

Noh and Park 
introduced an  extra free parameter $r$ controlling memory. 
In this model, if pair of occupied sites is chosen, it can either  
lead to creation of one more particle with some probability, 
or both particles can be annihilated with certain probability. 
If this was the only dynamics, the state with no pairs is an absorbing 
state and such a transition is known to be in DP class. However, picture 
changes if we allow particles to diffuse. 
For nonzero diffusion, we has two absorbing states. Apart from vacuum 
state, we have a state with single diffusing particle. Such  state 
is not unique. It can be termed as 
$L$ dimensional absorbing subspace to which particle is confined. 

The pair density in this model can change due to diffusion which
leads to long-term memory. The pair created at some time due to diffusion
 can proliferate and lead to extra particle after a very long time.
The parameter $r$  controls creation of
diffusion-induced pairs. The simulation is carried out in 
following manner.
Consider a 1-d lattice of $N$ sites with periodic boundary conditions.
A pair of sites $i$ and $i+1$ is randomly selected.  If both sites 
occupied, the particles are
annihilated with probability $(1-d)p$ or a particle is added
to a  neighboring  $i-1$ or $i+2$'th  site with probability $(1-d)(1-p)$
provided neighboring site is empty.
If only one of the $i$ and $i+1$'th site  is occupied, 
the occupied particle hops to other site with probability $d$. 
If  hopping solitary particle creates a pair by coming in contact
with site occupied by a particle,
both particles are annihilated with probability
$1-r$ as mentioned before. One time-step is completed after $N$ 
such trials. In short, if diffusion leads to formation of particle pair,
such pair is
annihilated with probability $1-r$. For $r=0$, such pair is certainly
annihilated and diffusion does not increase number of particle pairs. 
For $r=0$, when the system reaches an absorbing state of PCP,
{\it{ i.e.}} there are no consecutive occupied sites, the dynamics
thereafter is just diffusion. If this diffusion leads to particle
pairs the particles are annihilated and the final absorbing state can be
a vaccum state or a solitary particle.
The number of absorbing states is not infinite for $r=0$ case
as it is for $d=0$ case.  But the absorbing state for $r=0$ case
is same as that for PCP since the dynamics thereafter is
plain diffusion.
In this case, it is reasonable
to expect that the transition is in DP class
\cite{lubeck-2,noh}. 
For $r=1$,  diffusion occurs independently and  can
change the number of particle pairs. This case is equivalent to PCPD. 
Noh and Park relate the additional parameter $r$ introduced by
them to memory since if 
diffusing particles can 
lead to particle pairs, pair-creation rate will be history dependent. 
This history dependence will not be there if the diffusing particles 
certainly annihilate on meeting. 

Noh and Park have reported accurate critical points at
which transition is observed
for  $d=0.1$. The values are $p_c=0.046872, 0.055055,
0.066364, 0.083155$ and $0.1112$ for  $r=0,0.25, 0.5, 0.75$ 
and $r=1$. For $r=1$ an improved value $0.111158$ has been reported by 
Park\cite{park-14}.
They 
investigated pair density and 
systematically found values of all
exponents for various values of $r$.
They found continuously varying exponent for various values of $r$.
We are not aware of any further 
theoretical or numerical studies in GPCPD after Noh and Park's work.

\section{Results}

We investigate this problem using persistence as `order parameter'.
We define persistent sites at time $T$ as ones which did not change their
initial state at all time steps $t\leq T$.
We find that the asymptotic persistence is zero for $p<p_c$.
The $p_c$  values at which persistence goes to zero 
match with $p_c$  values mentioned above.
We obtain excellent power-laws for our order parameter decay 
for GPCPD at these  points.

We have simulated the  GPCPD model for lattice size of $N=2 \times 10^5$ 
for $r=0, 0.25, 0.5, 0.75$ and $r=1$ and  $d=0.1$ and average over
more than $3-5 \times 10^5$ configurations.
In Fig. 1a), we plot fraction of persistent sites, $P(t)$ as a function
of time $t$ for these values of $r$ at the critical point which 
exhibits a clear power-law decay. 
The exponent obtained ranges between 
1.935-2.22 and it increases with $r$.
If $P(t)\sim t^{-\theta}$, $P(t)t^{\theta}$ should be a constant. We have 
also plotted
$P(t)t^{\theta}$ as a function of $t$ in Fig. 1b), 
and it shows a flat line asymptotically.
 We observe a flat
line over several decades for $r \ne 1$. Unfortunately, for 
$r=1$ the onset of scaling is late and it is observed over fewer
decades.

We use  well established phenomenological scaling laws extended 
to persistence\cite{fuchs,ashwini}.  For system size $N$ and 
the distance from criticality $\Delta = \vert p-p_c \vert$, 
We expect following asymptotic law to hold
\begin{equation}
P_N(t) =t^{-\theta} F(t/N^z, t \Delta^{\nu_\parallel} )
\end{equation}
where F is the scaling function and $z=\frac{\nu_\parallel}{\nu_\bot}$ is 
dynamic scaling exponent. Though $\theta$ is not an universal exponent,
$z$ and $\nu_\parallel$ are. We hope that these scaling 
relations will shed light on possible
universality class of underlying models.

\begin{center}
\begin{figure}
\includegraphics[width=80mm]{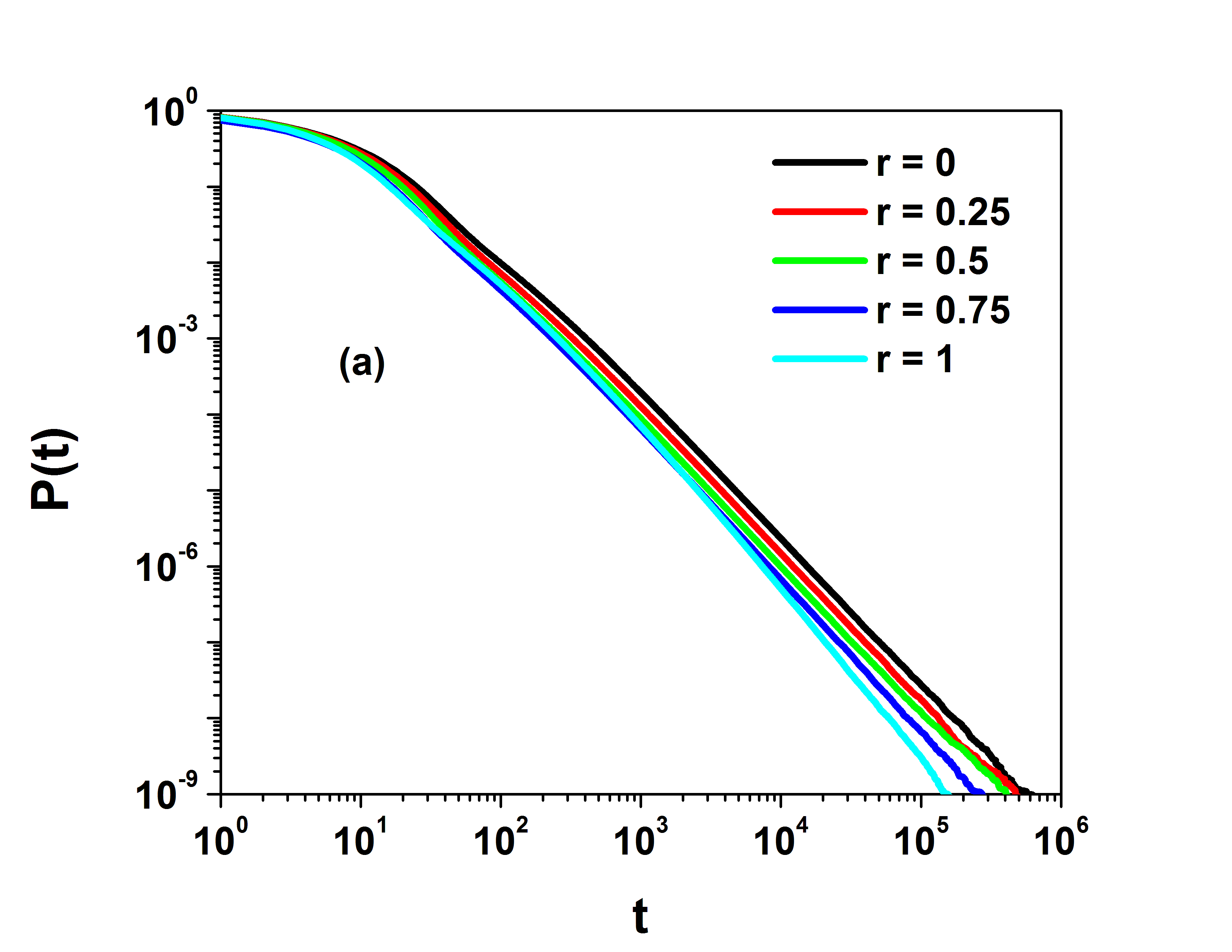}
\includegraphics[width=80mm]{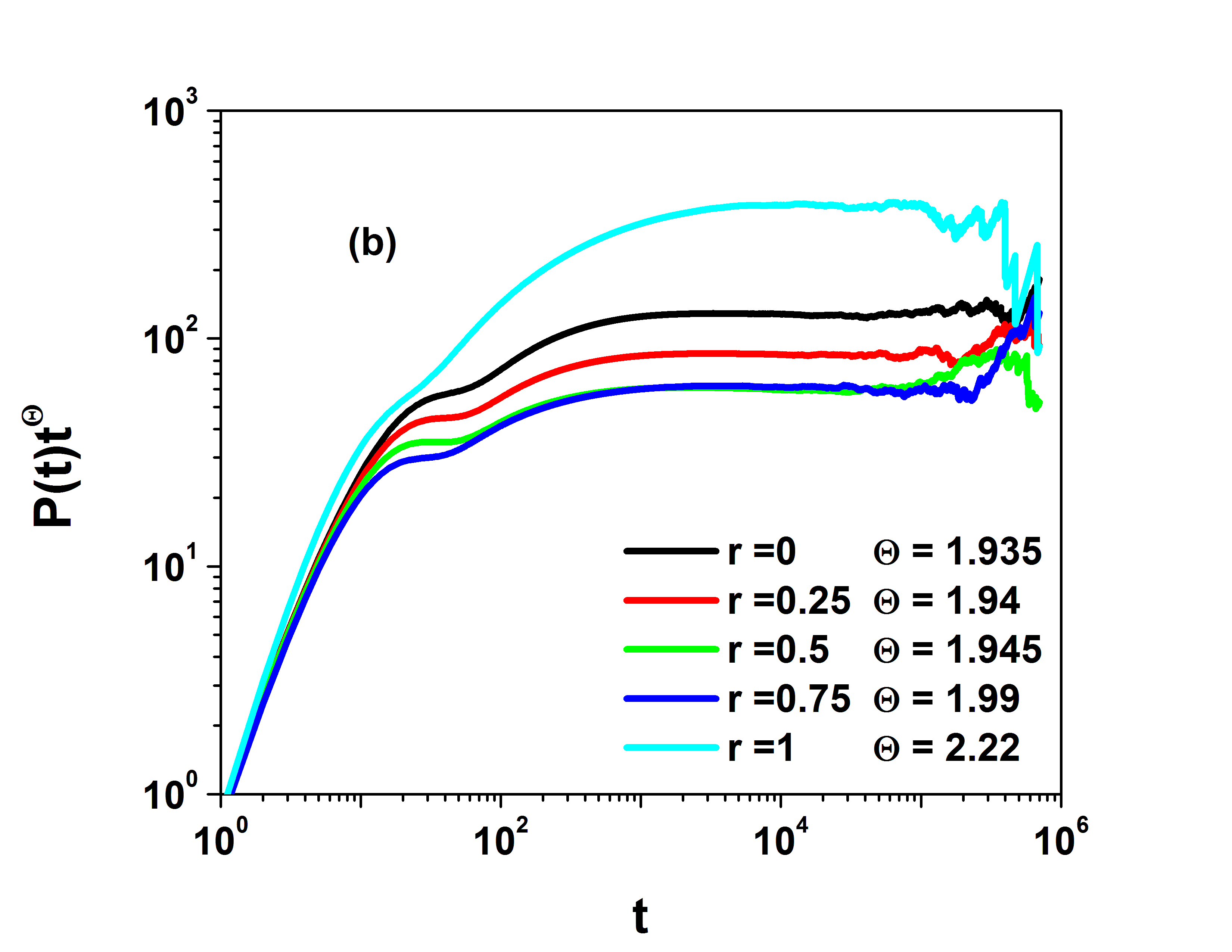}
\caption{a)$P(t)$ as a function of time t, for various values of $r$
 at critical point. We carry out simulation for  N=$3-5 \times 10^5$ 
sites and average over $2\times 10^5$  configurations.
b) $P(t)t^{\theta}$ as a 
function of time $t$, for various values of $r$ at the critical point for
 same data as (a).}
\label{figure 1}
\end{figure}
\end{center}

\begin{figure}[ht!]
\begin{center}$
\begin{array}{c}
\centering
\includegraphics[width=95mm]{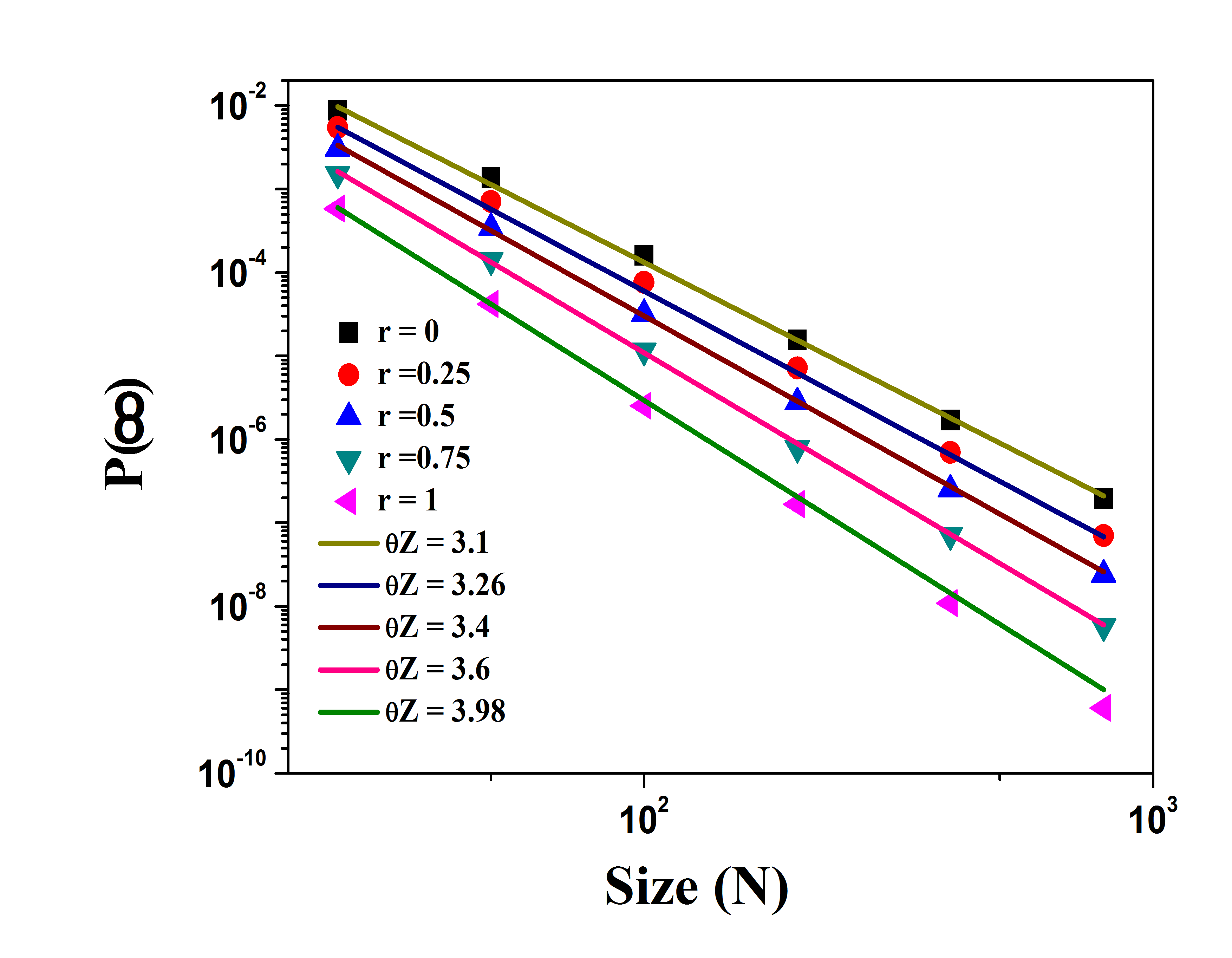}
\end{array}$
\end{center}
\caption{a) Saturation value of Persistence $P(\infty\})$  as a function 
of $N$ is plotted for various value of $r$ at $p=p_c$. 
We wait for $10^8-10^{10}$
time-steps and average over $6\times 10^5 - 10^7$ initial 
conditions with longer waiting period  and higher averaging for larger $N$.}
\label{figure 2}
\end{figure}

\begin{figure}[ht!]
\begin{center}$
\begin{array}{cc}
\includegraphics[width=80mm]{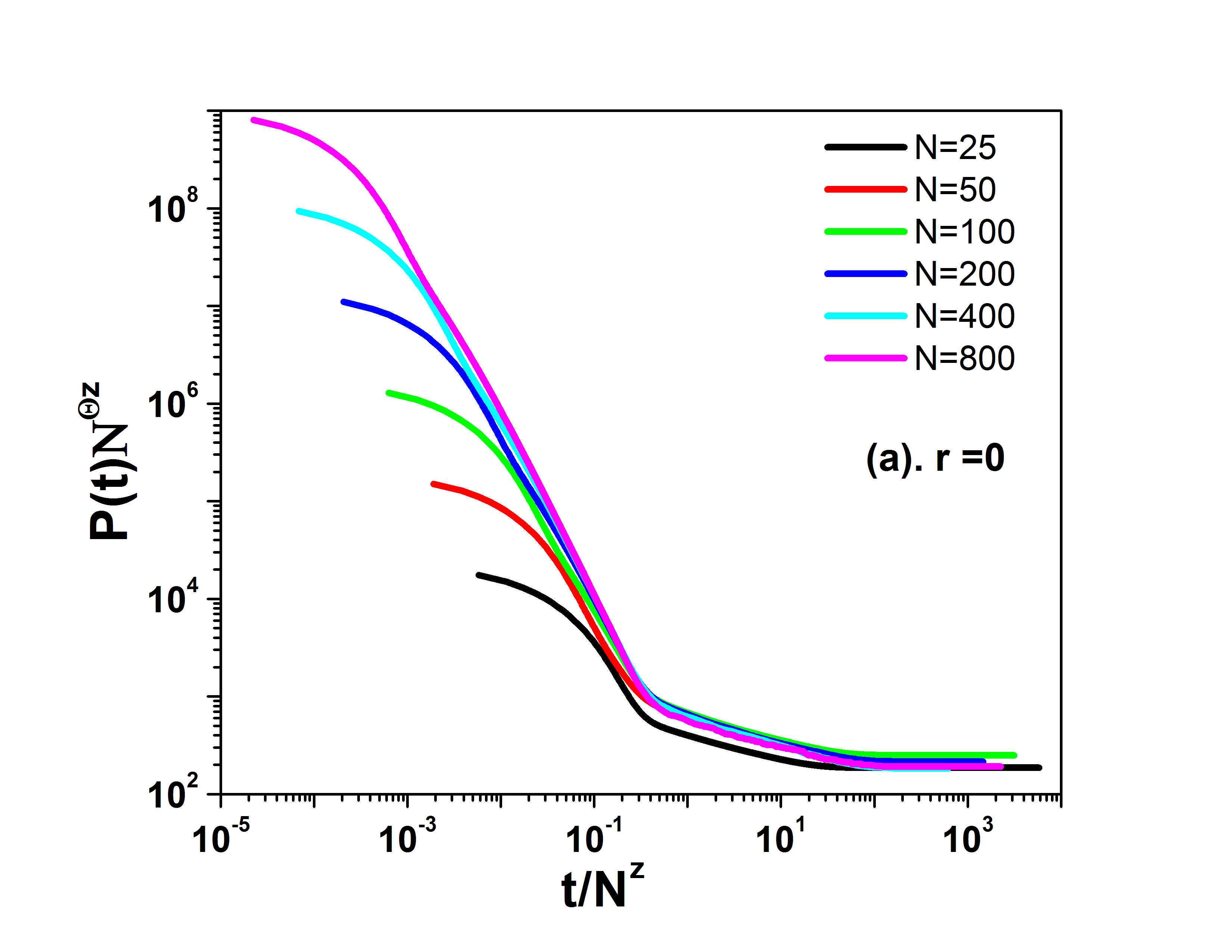} &
\includegraphics[width=80mm]{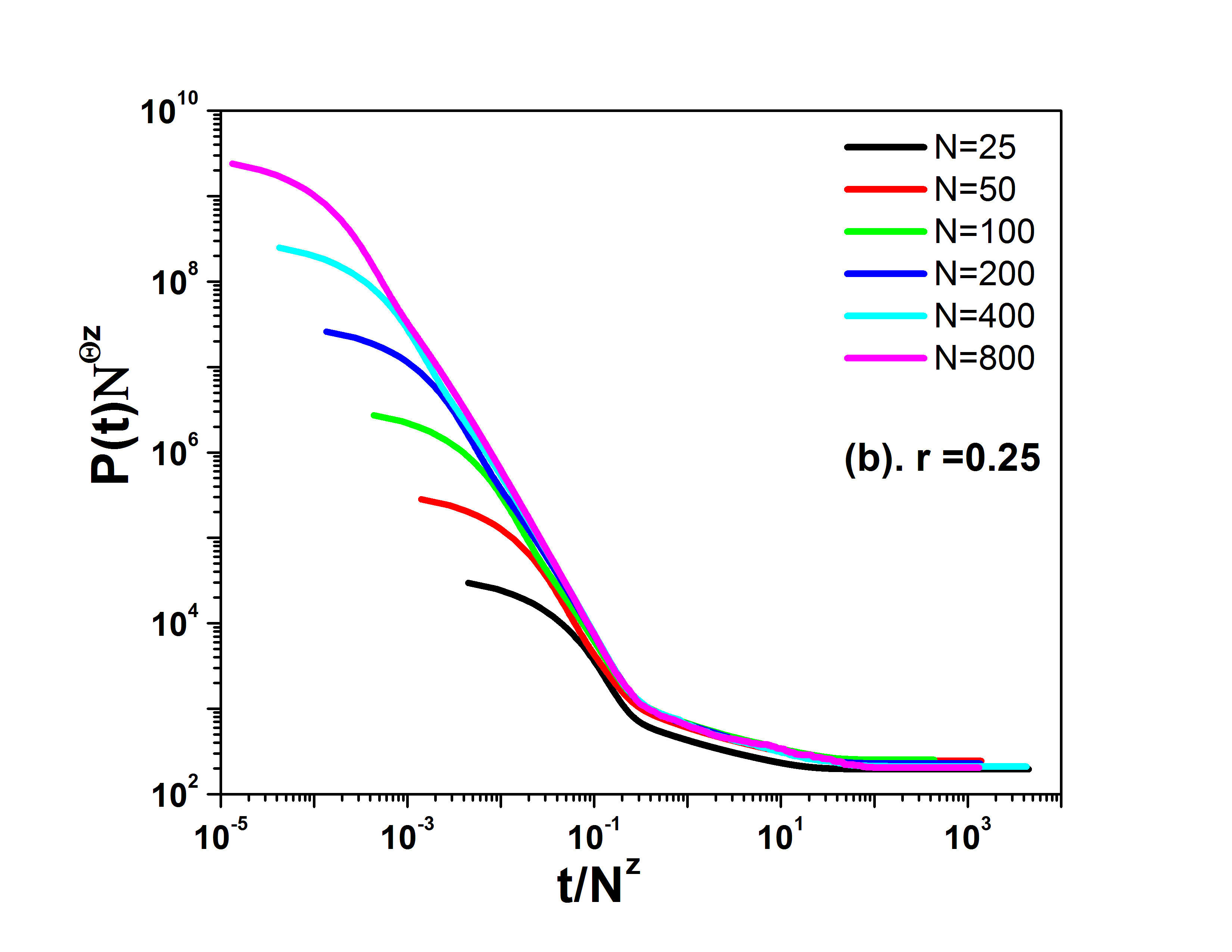}\\
\includegraphics[width=80mm]{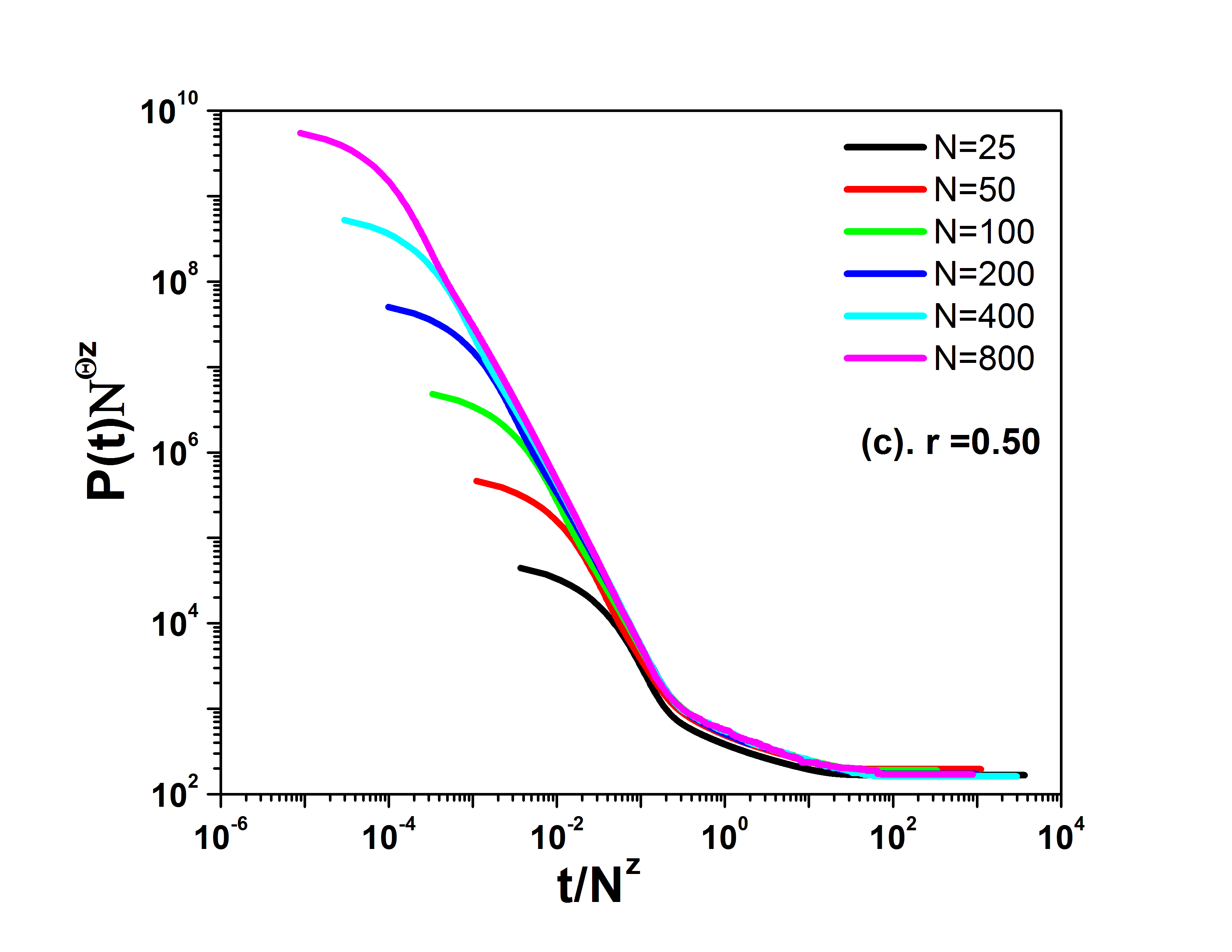} &
\includegraphics[width=80mm]{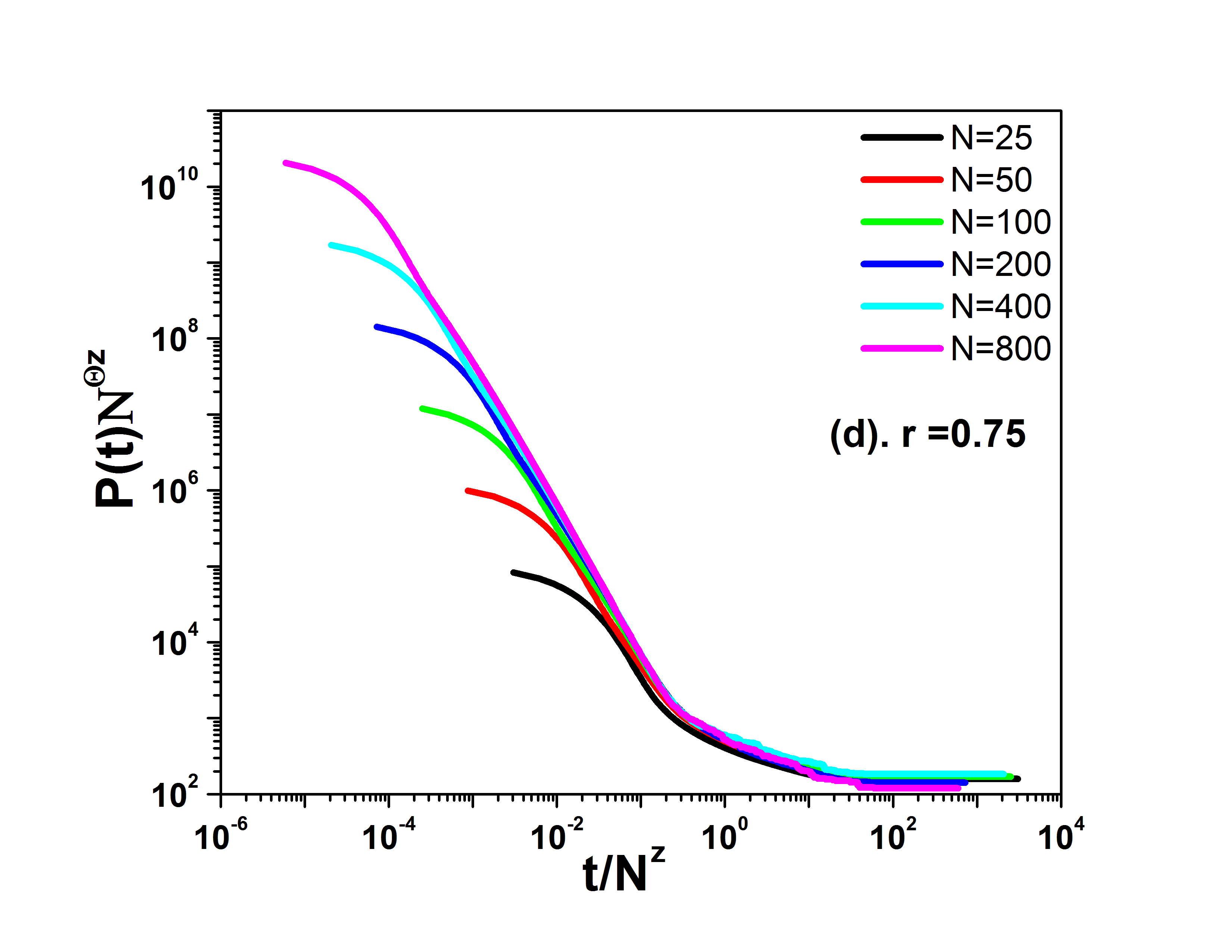}\\
\includegraphics[width=80mm]{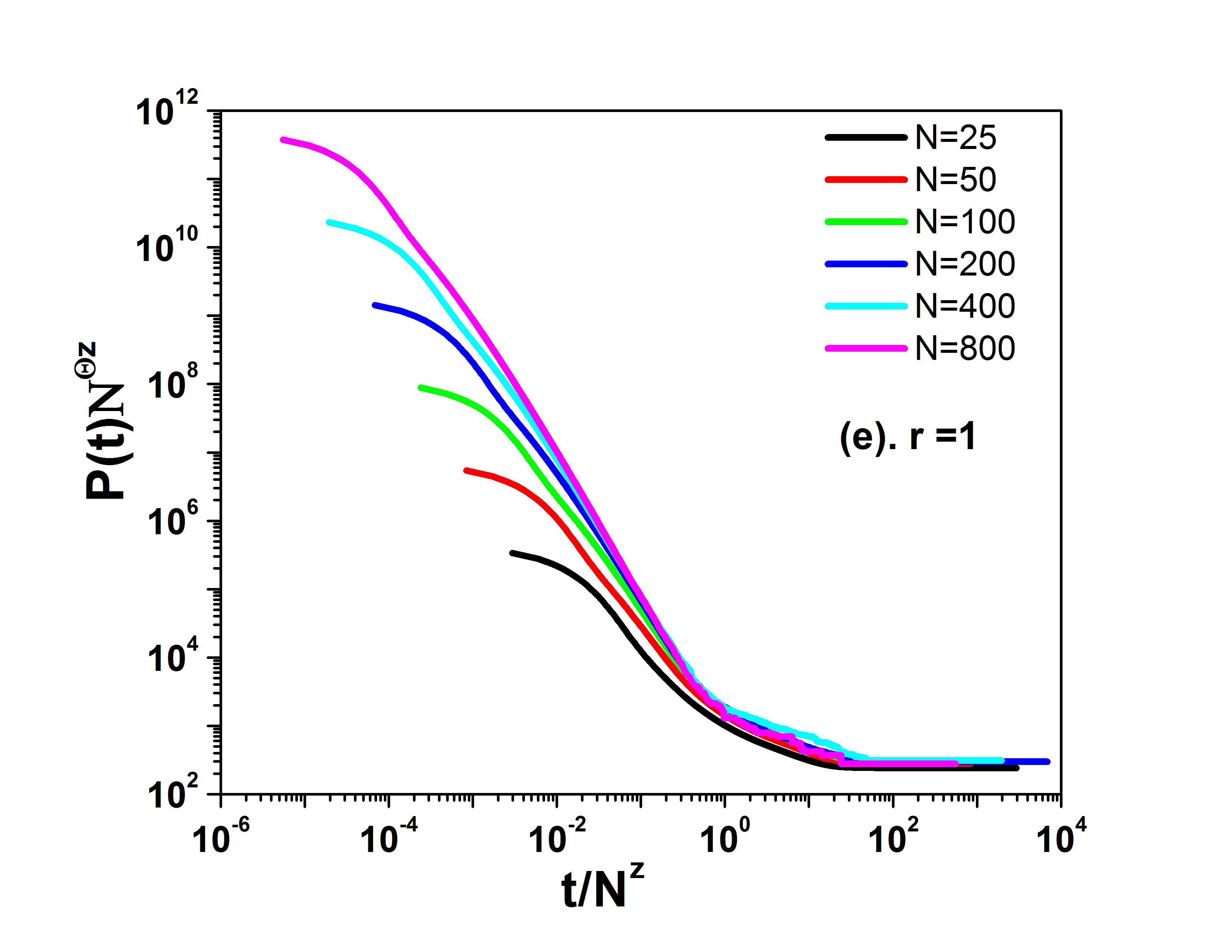} &
\end{array}$
\end{center}
\caption{Persistence $P(t) N^{z\theta}$ as a function of $t/N^z$ for 
various values of $N$ at critical point $p=p_c$. We 
average over $10^5-10^6$ configurations with more averaging 
for larger values of $N$.}
\label{figure 3}
\end{figure}

\
\begin{figure}[ht!]
\begin{center}$
\begin{array}{cc}
\includegraphics[width=80mm]{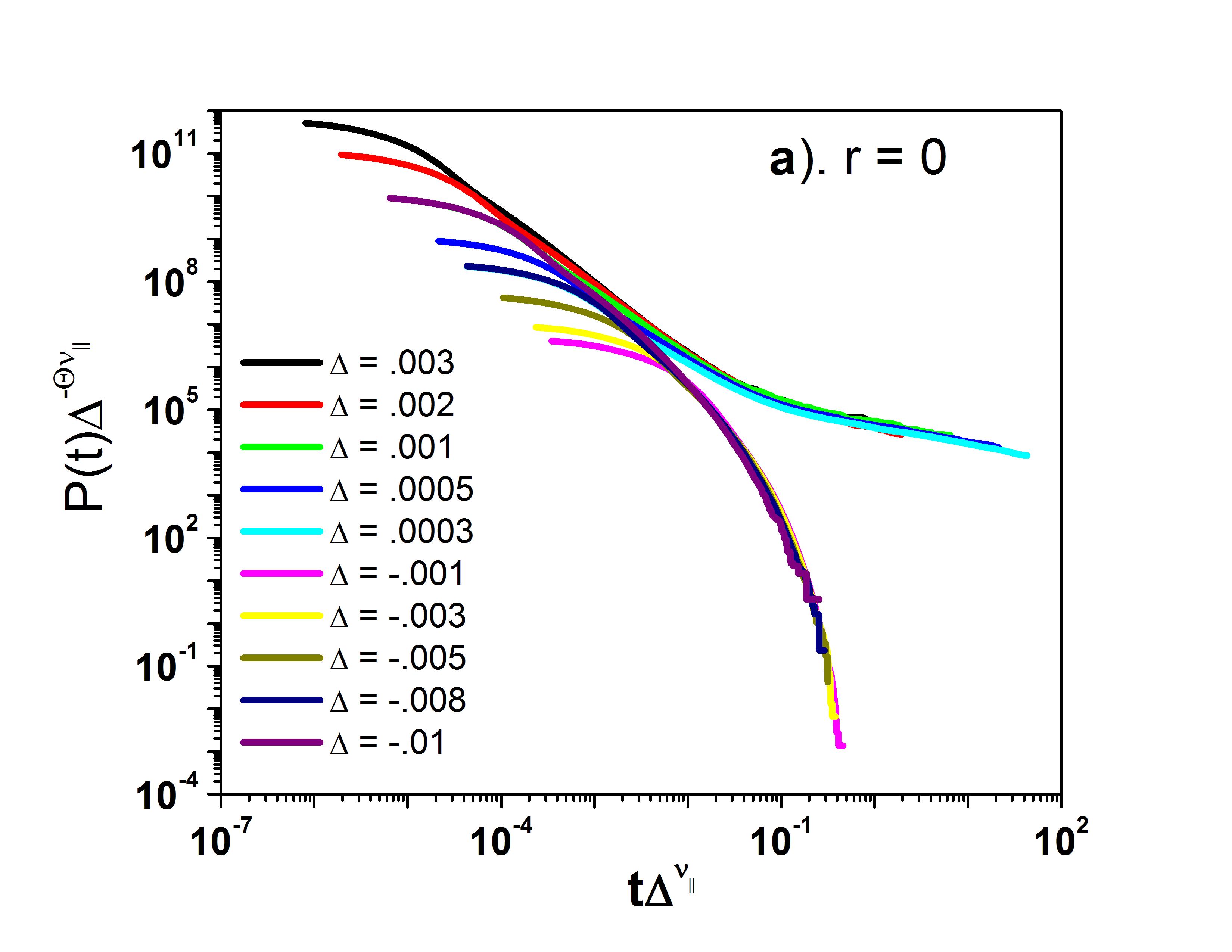} &
\includegraphics[width=80mm]{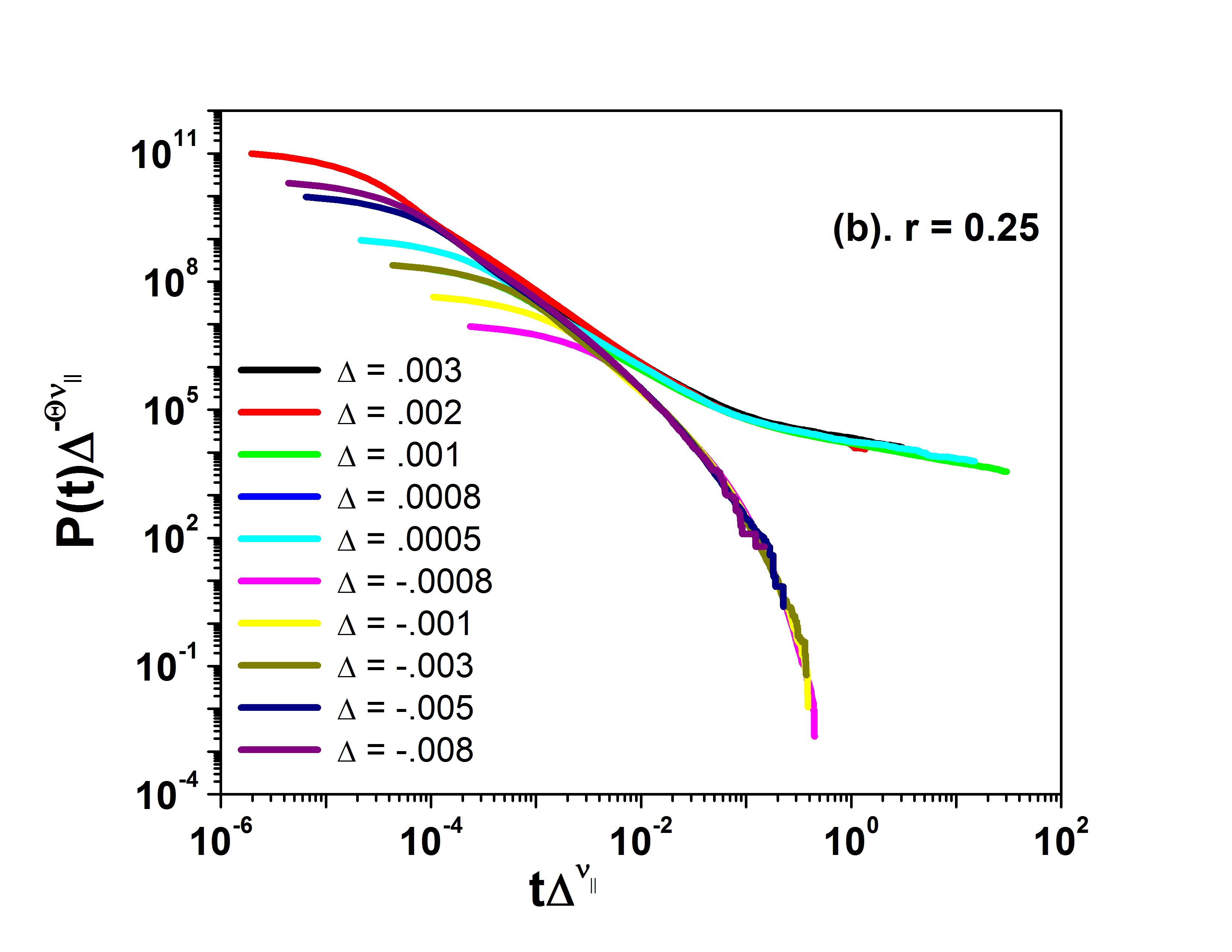}\\
\includegraphics[width=80mm]{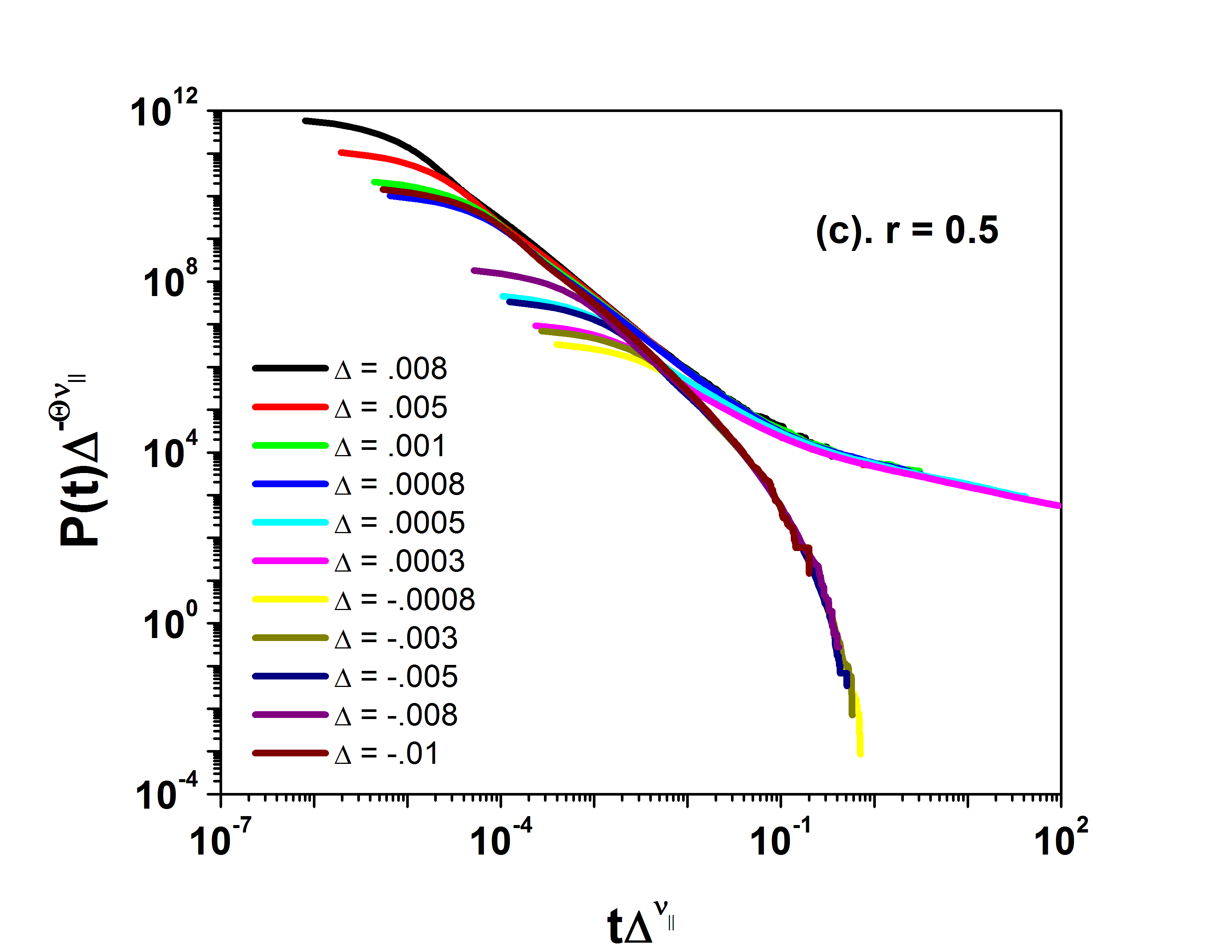} &
\includegraphics[width=80mm]{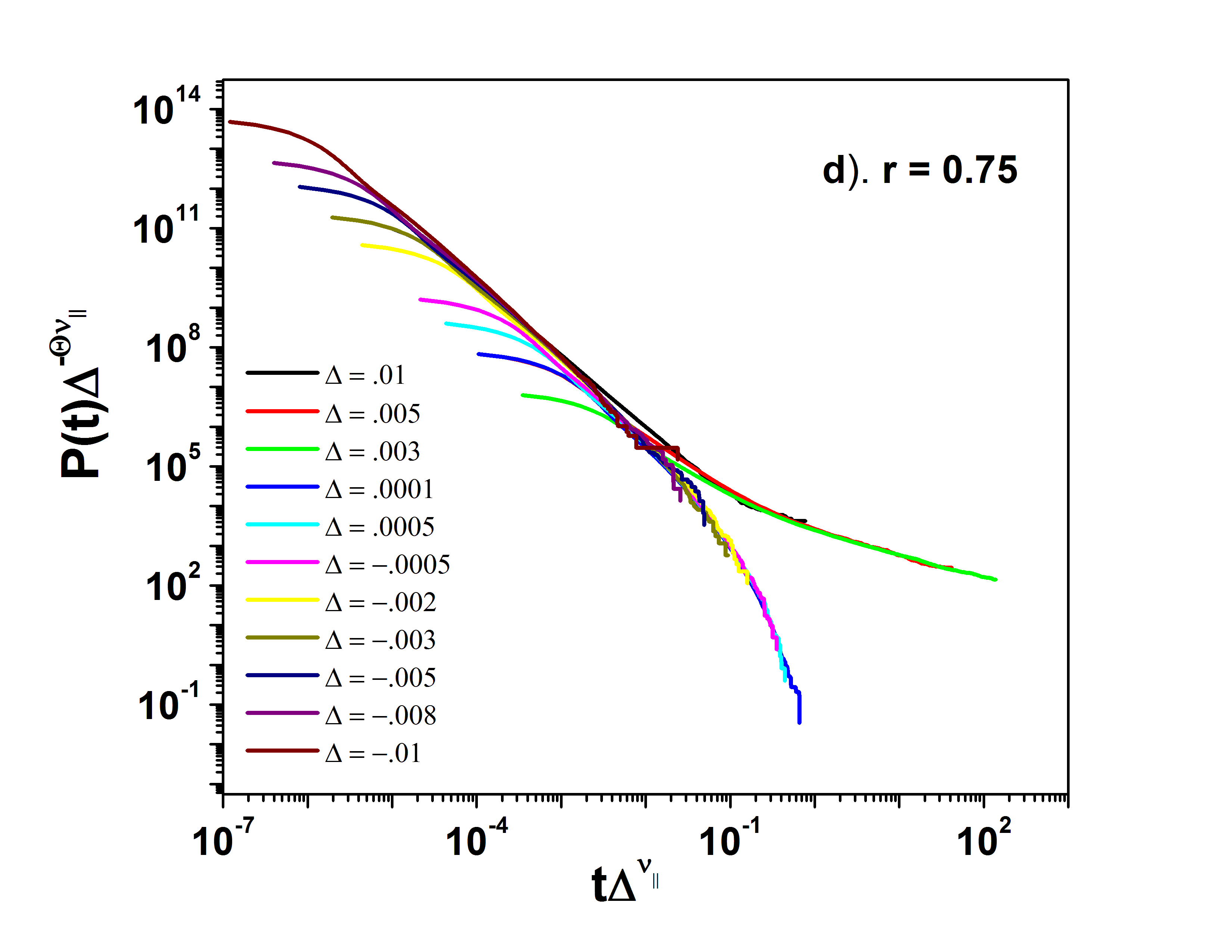}\\
\includegraphics[width=80mm]{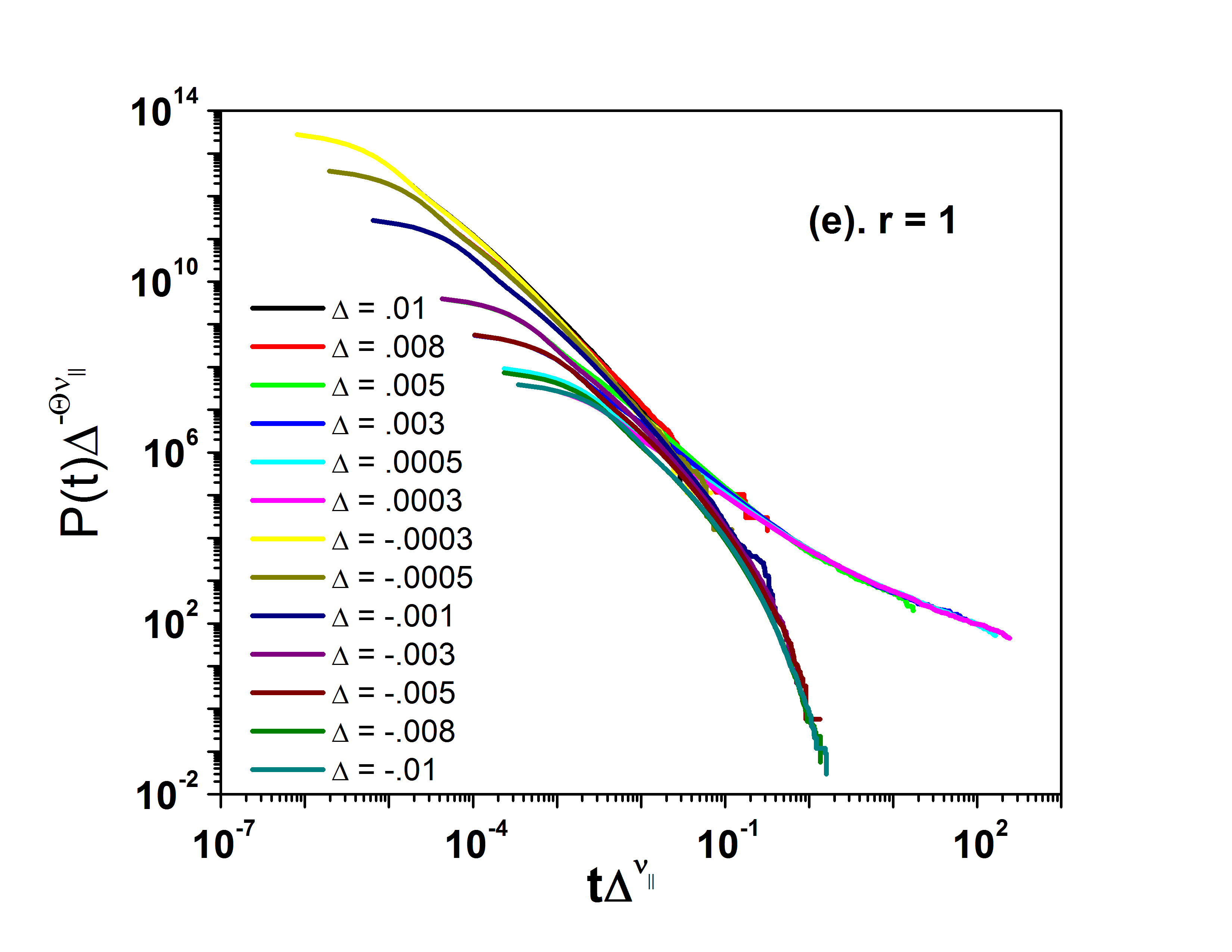} &
\end{array}$
\end{center}
\caption{Persistence $P(t)\Delta^{-\theta\nu_\parallel}$ 
is plotted as a function of
$t\Delta^{\nu_\parallel}$ 
for various values of $\Delta=\vert p-p_c\vert$. 
The limiting curves are different for $p>p_c$ and $p<p_c$. We average over
$5\times 10^4$ configurations for $p<p_c$ and $5\times 10^3-10^4$ configurations
for $p>p_c$.}
\label{figure 4}
\end{figure}

We consider two special cases.

a) Finite-size scaling : 
We find the persistence probability 
at the critical points $p=p_c$,  for 
various values of $r$ for 
finite sizes.  We define $\Delta=\vert  p-p_c \vert$. Thus
we conduct simulations for $\Delta=0$ for various values 
of $r$. Thus we expect, $P_N(t) \simeq  t^{-\theta} F(\frac{t}{N^z})$.

Like order parameter, the persistence 
probability saturates to  some value for $t>N^z$. The saturation value
for $P(t)$ is expected to be proportional to $N^{-z\theta}$. Thus
plotting $P(t) N^{z\theta}$ as a function of $t/N^z$ is expected
to yield a good scaling behavior.
In Fig. 2, we have plotted saturation value of persistence $P(\infty)$ at
critical point for various values of $N$. The slope is
expected to yield value of $z\theta$ which increases monotonically with
$r$ and the values of $z$ are not compatible with DP for these sizes.

We have plotted $P(t) N^{z\theta}$ as a function of $t/N^z$ 
at $p=p_c$ for various values of $r$ shown in Fig. 3. 
A good scaling behavior is
obtained for values of $z$ obtained from $z\theta$ values 
inferred from saturation value of persistence in Fig.2. 
A successful scaling
collapse  is obtained for all values of $r$.
The value of $z=1.60$ for
$r=0$ is close to  DP exponent $z=1.581$.
For $r>0$, our values  of $z$ increase with $r$  as reported in \cite{noh}.
The precise values are slightly larger than those reported above. 
The value $z$ for $r=1$ is  
in agreement Noh and Park's estimate from static simulations and
there is no report of $z<1.8$  from static simulations in earlier works.
This value is also
in agreement with recent estimates from
static simulations without assuming any corrections to scaling. (See table
II of \cite{smallenberg}) 
The  values of $z$  reported in literature decrease with $d$ and approach DP
as $d\rightarrow 1$(See table III of \cite{henkel}). 

The value of $\theta$ changes slowly for smaller values of $r$ and
shows a big jump at $r=1$. This change is not reflected in $z$ which 
saturates for large value of $z$.

Obtaining good statistics for saturation values of $P(t)$ for larger $N$ is 
made difficult due to the fact that $z\theta>>1$. If $z\theta=1$(say), and 
4$\%$ sites were persistent for $N=100$, $1\%$ sites will be persistent for 
$N=400$ and number of persistent sites for $N=400$ will be 4 again. The 
expected {\em {number}} of persistent sites is actually same for single 
configuration.  It is easy to see that the number of configurations required 
to have reliable statistics grows like  $N^{z\theta-1}$. If $z\theta <1$,  
less averaging is required for large $N$, For $z\theta>1$,  a single 
configuration  of with large $N$ will have no persistent site at critical 
point asymptotically.  Apart from the fact that we have to wait for longer 
time for larger values of $N$, much higher averaging is required to obtain 
reliable statistics.  Hence we have not carried simulations for larger sizes.
Another reason is that uncertainty in critical point would lead to significant 
errors in value of $z$ for large $N$.  Overestimate of $p_c$ will lead to 
underestimate of value of $z$ and vice versa.

b) Off-critical simulations: We conduct simulations for 
$N=2\times 10^5$ and average over at least $8\times 10^3$ 
initial conditions for 
$p<p_c$ and at least 
$10^3$ initial conditions for $p>p_c$. The value of $z\theta$ 
ranges from $3.1-3.98$ and saturation value at critical point for $N=10^4$ 
will be of order $N^{-z\theta}\sim10^{-12}$ or less  
and below numerical precision.
For all practical purposes, finite size corrections can be ignored. 
The scaling form mentioned above indicates a standard scaling collapse
when we plot $P(t)\Delta^{-\theta\nu_\parallel}$ as
a function of $t\Delta^{\nu_\parallel}$ where $\Delta=\vert p-p_c
\vert$.   The persistence saturates for $p>p_c$ and 
goes to zero for $p<p_c$ which is exactly opposite to the behavior
of particle density or pair density.
We carry out this scaling for various values of $r$ and find 
value of $\nu_{\parallel}$
In Fig. 4, we have shown scaling collapse for various
values of $r$ and $\Delta$.The best fitting values of 
$\nu_\parallel$ are given in table  below.
For $\nu_\parallel=1.73$, we obtain excellent scaling
collapse for $r=0,0.25,0.50, 0.75$ and $1$. This value 
of $\nu_\parallel$ matches with that of DP.

\begin{table}
\caption{Critical exponents for GPCPD and comparison with
Ref. \cite{noh}}
 \label{tab1} 
\begin{tabular}{|c|c|c|c|c|c|c|}
\hline
 r  & $\nu_\parallel$\cite{noh} & $\nu_\parallel$ &  $z$\cite{noh} &  $z$ &$\theta$ \\
 \hline
 0    &  1.738 &1.73  & 1.58 & 1.60 & 1.935 \\ 
 \hline
 0.25 &   1.80  &1.73   & 1.64 & 1.68 & 1.94 \\
 \hline
 0.5  &   1.86  &1.73 & 1.69 & 1.74 & 1.945\\
 \hline
 0.75 &   2.01  &1.73  & 1.72 & 1.80 & 1.99\\
 \hline
 1    &   2.34  & 1.73  & 1.8  & 1.81 & 2.22 \\ 
\hline
\end{tabular}
\end{table}

\begin{table}
\caption{  Comparison of DP,  PCPD and DI}
\begin{tabular}{|c|c|c|}
\hline
 Class  &  $\nu_\parallel$ &z\\  
 \hline
 DP \cite{noh}  & 1.73 & 1.58 \\
 \hline
 PCPD (d=0.1)  & 1.61- & 1.8-\\
\cite{henkel,smallenberg}  & 2.45 & 2.08 \\
 \hline
 DI \cite{noh}  &  3.22 & 1.75 \\
\hline
\end{tabular}
\end{table}

\section{Discussion}

We have found that in GPCPD 
a)Persistence decays as power law at critical point.
b)The persistence exponent is a function of $r$ and 
c)The persistence
exponent is much larger than ${\delta}$ and presents a better 
scaling behaviour.  Thus
it can be used to find other exponents such as $z$ and 
$\nu_\parallel$. Of course, all 
three critical exponents cannot be found by this 
approach (since ${\delta} \neq \theta$).
However, it is not necessary that there will be a well 
defined persistence exponent
at the critical point, or that the persistence exponent 
will be larger than ${\delta}$.
Only argument we can give for validity of phenomenological 
scaling laws is cases
in literature where it has been successfully used. 
It is necessary to study 
the scaling of persistence and validity of various scaling laws in cases 
where the university class of model is well established.
This will enable this  
approach is placed on a firm footing, at least numerically.

Our values of $z$ are broadly in agreement with Noh and Park. 
We have not been able to go for very large sizes for computation of
$z$ since the saturation value of persistence even for $N=400$ or $800$ is 
reached only after $10^9$ timesteps or more. This makes it very difficult
to go to larger sizes.
It is very unlikely that the static exponent $\nu_\perp$ changes with $r$ and
$\nu_\parallel$ does not change.  
We believe that simulations for larger sizes and longer times will yield 
value of $z$ which is compatible with DP.

Noh and Park speculate, that their model has continuously varying set of 
exponents.  Given the broad range of exponents obtained in literature for 
PCPD, there was a speculation that
PCPD has continuously varying set of
exponents\cite{henkel}.
In  Table 3 of ref. \cite{henkel}, several 
values of critical exponents for 1D PCPD  for $d=0.1$ from
literature are listed.
The values of $z$ range from 1.8 \cite{noh} to 2.04 \cite{dickman}. 
The values of $\nu_\parallel$ can be obtained   $\nu_\parallel=z \nu_\perp 
= \frac{z\beta}{\frac{\beta}{\nu_\perp}}$
and they range from 1.61 \cite{odor-2} to 2.45\cite{hinrich-01}. 
Needless to say, there is even further variation with $d$ and
authors generally report smaller values of $z$  for larger $d$.
It has been proposed that PCPD has continuously varying set of
exponents depending on value of $d$\cite{dickman}.
Noh and Park conjectured that it 
is possible that GPCPD is in same class with further variation 
due to extra free parameter $r$.  However,
the idea of continuously varying exponents is not very compatible
with classical idea of universality which leads to few exponents depending
on dimensionality and symmetries\cite{gredat}. 
Recent works in PCPD have suggested
a clear drift of exponents toward DP values for long-time simulations.

Our work suggests that the value of $\nu_\parallel$ in GPCPD is same
as in DP. However, our value of $z$ differs considerably from DP values.
This is a feature shared by analysis of all static simulations in PCPD
without assuming power-law or logarithmic correction.
This departure from DP values could be an artifact of finite size and
finite time simulations. In a cellular automata model analogous to PCPD,
Hinrichsen has clearly demonstrated that the values of $z$ reduce as
a function of $t$\cite{hinrich2}
and static simulations may not be the best way to 
obtain dynamic exponent $z$.

We have revisted the problem of universality class of PCPD
with an alternative approach. 
Despite extensive numerical effort, we are not able to 
completely settle this  issue. However, above results indicate
that we cannot rule out the possibility that
PCPD is in same universality class as DP.

\section*{Acknowledgements}
MBM would like to thank UGC-BSR fellowship.  PMG 
would like to thank RTMNU research project for financial assistance.
PMG would like to thank P. K. Mohanty and P. Sen for useful discussions
and VNIT for allowing the use of their computational facility..

\section*{References}

\begin{thebibliography}{}
\bibitem{latorre} Grassberger P and  de la Torre A, 1979 Annals of Physics
{\bf {122}}, 373. 
\bibitem{grassberger} Janssen H K, 1981
Z. Phys.  B {\bf{42}} 151, Grassberger P, 1982
Z. Phys.  B {\bf{47}} 365.
\bibitem{jensen}Jensen I, 1994 Phys. Rev. E  {\bf{50}}, 3623.
\bibitem{grassberger-ca}Grassberger P, 1989 J. Phys. A {\bf {22}}, L1103.
\bibitem{park} Park H,  Kim M H and Park H, 1995 Phys. Rev. E
{\bf{52}}, 5664.
\bibitem{odor}
 Menyh\'ard N, 1994 J. Phys. A {\bf{27}}, 6139;
Menyh\'ard N and 
\'Odor 
G, 1996 J. Phys. A {\bf{29}}, 7739.
\bibitem{park-3}Hwang W, Kwon S,  Park H, and  Park H, 1998 Phys. Rev. E,
{\bf{57}}, 6438.
\bibitem{park-park} Park, S C and Park H, 2008  
Eur. Phys. J. B, {\bf{64}}, 415.
\bibitem{Mendes} See {\it{e.g}}
Mendes J F F,  Dickman R and Herrmann H, 1996 
Phys. Rev. E, {\bf {54}} R3071(R).
\bibitem{mohanty}  Basu M,  Basu U, Bondyopadhyay S,  Mohanty P K and
 Hinrichsen H, 2012 Phys. Rev. Lett. {\bf 109} 015702,
Lee S B, 2014 Phys. Rev. E,  {\bf 89} 060101.
\bibitem{odor-rev} \'Odor G 2004 Rev. Mod. Phys. {\bf{76}}, 663.
\bibitem{lubeck-2}
L\={u}beck S and Willman R D, 2002 J. Phys. A {\bf {35}},
10205,  Jensen I, 1993  Phys. Rev. Lett. {\bf 70}, 1465,
Jensen I and Dickman R, 1993 Phys. Rev. E 
{\bf 48}, 1710.
\bibitem{tcpd}
Further generalization
of this model such as TCPD (Triplet contact process with diffusion) is
also studied , see {\it {e.g.}}
Schram R D and Barkema G T, 2013 J. Stat. Mech. P04020.
\bibitem{hinrich-01}Hinrichsen H, 2001
Phys. Rev. E {\bf {63}} 036102.
\bibitem{henkel} Henkel M and Hinrichsen H, 2004, J. Phys. A: Math. Gen.
{\bf 37}, R117.  
\bibitem{physica}
Hinrichsen H, 2006 Physica A {\bf{361}}, 457.
\bibitem{park-14}
 Park S-C, 2014 Phys. Rev. E {\bf {90}}, 052115.
\bibitem{schram}
Schram R. D. and Barkema G. T, J. Stat. Mech. 2012 P03009.
\bibitem{smallenberg}
 Smallenburg F and  Barkema G T, 2008, Phys. Rev. E {\bf{78}}, 031129
\bibitem{gredat}  See, for example,
 Gredat D {\it{et al}},  2014 Phys. Rev. E {\bf 89} 010102.
\bibitem{Janssen}
Janssen H K, Schaub B, and Schmittmann B, 1989 Z. Phys. B: Condens. Matter 
{\bf 73}, 539.
\bibitem{Huse}
Huse D A, 1989  Phys. Rev. B {\bf 40}, 304.
\bibitem{majumdar}For a review, see  Majumdar S N, 1999, Curr. Sci.
{\bf 77}, 370.
\bibitem{Zheng} See, e.g., 
Zheng B, 1998 Int. J. Mod. Phys. B {\bf 12}, 1419,
 Arashiro E,  Drugowich de Felicio J R, 2003 Phys. Rev. E 67 046123.
\bibitem{fuchs} Fuchs J,   Schelter J, Ginelli F and Hinrichsen H, 2008
J. Stat. Mech. P04015
\bibitem{footnote2}In spin systems, a spin can flip even if 
it is surrounded by neighbors having same spin value at any finite 
temperature. Thus persistence decays
exponentially at all finite temperatures.
In contact processes, uninfected site will not be
infected unless it has an infected neighbor.
\bibitem{stauffer}
 Stauffer D and de Oliveira P M C, 2002 Eur. Phys. J. B {\bf 30}, 587.
\bibitem{ggs}
Gade P M and  Sahasrabudhe G G, 2013
Phys. Rev. E,  {\bf 87}, 052905.
\bibitem{koduvely}
 Hinrichsen H and  Koduvely H M, 1998 Eur. Phys. J. B {\bf 5} 257.
\bibitem{albano}Albano E V and  Munoz M A, 2001 Phys. Rev. E
{\bf 63} 031104.
\bibitem{grass} Grassberger P, 2009  J. Stat. Mech. P08021
\bibitem{menon} Menon G I,  Sinha S  and  Ray P, 2003 Europhys. Lett.
{\bf 61} 27.
\bibitem{ali-gade} Saif M A and  Gade P M, 2010
J. Stat.  Mech. :Theory and Experiment P03016.
\bibitem{manoj}
Manoj G, and  Ray P, 2000 Phys. Rev. E {\bf 62}, 7755.
\bibitem{noh} 
Noh J D and Park H, 2004 Phys. Rev. E {\bf 69} 016122.
\bibitem{ashwini}Mahajan A V and  Gade P M, 2010
Phys. Rev. E {\bf 81}, 056211. 
\bibitem{dickman}  Dickman R and  de Menzes M A F, 2002
 Phys. Rev. E {\bf
66}, 045101.
\bibitem{odor-2}
\'Odor 
G, 2003 Phys. Rev. E,  {\bf 67} 016111.
\bibitem{hinrich2}
H. Hinrichsen, Physica A, 2003 {\bf {320}}, 249.
\end{thebibliography}

\end{document}